\documentclass[sigconf, authorversion=true]{acmart}

\copyrightyear{2024}
\acmYear{2024}
\setcopyright{acmlicensed}\acmConference[DAC '24]{61st ACM/IEEE Design Automation Conference}{June 23--27, 2024}{San Francisco, CA, USA}
\acmBooktitle{61st ACM/IEEE Design Automation Conference (DAC '24), June 23--27, 2024, San Francisco, CA, USA}
\acmDOI{10.1145/3649329.3656541}
\acmISBN{979-8-4007-0601-1/24/06}

\usepackage{algorithm}

\usepackage[noend]{algpseudocode}

\usepackage{etoolbox}
\usepackage[normalem]{ulem}

\PassOptionsToPackage{hyphens}{url}
\usepackage{url}

\usepackage{amsthm}

\theoremstyle{definition}
\newtheorem{definition}{Definition}
\newtheorem{observation}{Observation}

\newtoggle{STRIKEMODE}
\togglefalse{STRIKEMODE}
\newcommand{\CITE}[1]{\iftoggle{STRIKEMODE}{\textbackslash CITE\{#1\}}{\cite{#1}}}

\newcommand{\refalg}[1]{Alg.~\ref{alg:#1}}

\newcommand{\RISCV}{\mbox{RISC-V}}

\begin{document}

\title{%
MCU-Wide Timing Side Channels and Their Detection
}

\vspace{-1mm}
\thanks{This work was supported partly by Bundesministerium f\"ur Bildung und Forschung Scale4Edge, grant no. 16ME0122K-16ME0140+16ME0465, by Intel Corp., Scalable Assurance Program and by Siemens EDA}
\vspace{-3mm}

\author{Johannes M\"uller$^1$, Anna Lena Duque Ant\'{o}n$^1$, Lucas Deutschmann$^1$, Dino Mehmedagi\'{c}$^1$,\\ Cristiano Rodrigues$^2$, Daniel Oliveira$^2$, Keerthikumara Devarajegowda$^3$,\\ Mohammad Rahmani Fadiheh$^4$, Sandro Pinto$^2$, Dominik Stoffel$^1$, and Wolfgang Kunz$^1$}
\affiliation{
 	\institution{$^1$RPTU Kaiserslautern-Landau, Germany \, $^2$Universidade do Minho, Portugal \\$^3$Siemens EDA, Germany \, $^4$Stanford University, USA}
	\city{ }%
	\country{ }%
}

\renewcommand{\shortauthors}{J. M\"uller, A.L. Duque Ant\'{o}n, L. Deutschmann, D. Mehmedagi\'{c}, et al.}

\begin{abstract}
  Microarchitectural timing side channels have been thoroughly
  investigated as a security threat in hardware designs featuring
  shared buffers (e.g., caches) and/or parallelism between attacker
  and victim task execution. %
  However, contradicting common intuitions, recent activities
  demonstrate that this threat is real even in
  microcontroller SoCs without such features. %
  In this paper, we describe SoC-wide timing side channels previously
  neglected by security analysis and present a new formal method to
  close this gap. %
  In a case study on the RISC-V Pulpissimo SoC, our method
  detected a vulnerability to a previously unknown attack variant that allows
  an attacker to obtain information about a victim's memory access
  behavior. %
  After implementing a conservative fix, we were able to verify that
  the SoC is now secure w.r.t.\ the considered class of timing side channels. %
  \vspace{-5mm}
\end{abstract}


\maketitle

\section{Introduction}
\label{sec:introduction}

For the last three decades, microarchitectural timing side channels have been a subject of interest in academia and industry. %
Timing side channel attacks extract confidential data by measuring the time it takes to process data or by examining footprints that confidential data leaves in the system. %
While such attacks were demonstrated to exploit a wide range of microarchitectural features,
they mostly target systems that allow parallelism between attacker and victim execution or
systems featuring buffers such as caches that can hold a footprint of the confidential data. %

By contrast, microcontroller units (MCUs) are constructed with relatively simple
processor cores and usually offer neither attacker concurrency nor
footprint buffers. %
This may explain why they are neglected in the timing side channel literature. %

Recently, however, it has been shown that even simple MCUs can be vulnerable to timing side channel attacks~\CITE{2018-VanBulckPiessens.etal, 2022-BognarVanBulck.etal, 2024-RodriguesOliveira.etal}. %
The reported attacks exploit interrupts~\CITE{2018-VanBulckPiessens.etal} and memory bus arbitration~\CITE{2022-BognarVanBulck.etal,2024-RodriguesOliveira.etal} to bypass security infrastructure and steal confidential data. %
The BUSted attack~\CITE{2024-RodriguesOliveira.etal}, in particular, demonstrates a new class of timing side channels %
that exploit the plethora of complex IPs and on-chip communication structures (buses, interconnect)
that MCUs come with: %
An attacker task executing on the processor core accesses peripheral IPs
and instructs them to actively spy on a future execution of confidential programs by a victim. %
Instead of analyzing unintentional footprints left by the victim, the attacker
makes the spying IPs collect confidential information about the victim's memory accesses and save it in their own state.

In practice, the threat of vulnerable MCUs is of particular relevance
since such systems are found in virtually every industrial application domain. %
Apparently, a better understanding of SoC-wide timing side channels is needed, along with a
systematic approach to detect them. %
New formal methods must be developed capable of providing formal security
guarantees in MCUs. %
In case a side channel is detected, effective countermeasures are
required to reliably remove the vulnerability. %
This paper addresses these needs:
\begin{itemize}
\item We present, to the best of our knowledge, the first formal
  verification method to detect timing side channels in MCU-style
  SoCs. %
  The method is exhaustive w.r.t.\ the considered threat model. %
  We exploit critical characteristics of MCUs to
  formulate bounded properties spanning as little as two clock cycles
  and still obtain unbounded proof results, i.e., covering attacks that
  span potentially thousands of clock cycles. %
  (Sec.~\ref{sec:method})
\item With the proposed method, we discovered a vulnerability to a
  previously unknown variant of the BUSted~\CITE{2024-RodriguesOliveira.etal} attack. %
  In this scenario, an accelerator IP and a memory device can be
  leveraged to record and
  retrieve the memory access behavior of a victim task. %
  Compared to previous attack variants, no timer IP is required.
  (Sec.~\ref{sec:attack})
\item We conducted a case study on the Pulpissimo SoC~\CITE{2018-SchiavoneRossi.etal}. %
  We propose and implemented a countermeasure against timing side
  channels in SoCs and could prove that the system is secure. %
  (Sec.~\ref{sec:countermeasure})
\end{itemize}

\section{Timing side channels in SoCs}
\label{sec:model}

This paper focuses on timing side channels in MCU SoCs. %
In this section, we describe the threat model (Sec.~\ref{sec:threat-model})
and illustrate how timing side channels in SoCs can be exploited (Sec.~\ref{sec:attack-structure}). %

\subsection{Threat model}
\label{sec:threat-model}

We assume a system featuring a single-threaded single-core processor (simply called \emph{CPU} in this paper),
and several peripheral IPs and on-chip communication structures, as is commonplace in MCUs. %
The system supports multitasking by time multiplexing,
i.e., individual tasks do not execute simultaneously but are separated by a context switch. %
The security threat considered in this work consists of an \emph{attacker task}
and a \emph{victim task} executing on the CPU,
with the attacker trying to obtain confidential information about the victim execution. %
Our threat model addresses SoC-wide microarchitectural side channels.
Therefore, we can assume that the attacker has no direct access
to the memory regions allocated to the victim task and that
the victim does not store confidential data in attacker-accessible memory regions or IP registers.
Since the focus of this work is on MCU-style SoCs
without caches or special-purpose microarchitectural buffers, such as a branch history table,
we can further assume in our threat model that the confidential information does not leave footprints in the CPU.

\subsection{MCU timing side channels in a nutshell}
\label{sec:attack-structure}
  
A timing side channel attack on an MCU SoC usually consists of many steps~\CITE{2024-RodriguesOliveira.etal}.
For the purpose of this work, we concentrate on the key part of such attacks, i.e., we describe
how confidential information can be obtained by an attacker.
We refer the most interested readers to related work~\CITE{2024-RodriguesOliveira.etal}
for a complete understanding of the attack methodology.
In this work, we divide the exploitation of timing side channels in MCUs
into~3 phases which are separated by context switches: %
\begin{itemize}
\item \emph{Preparation:} The attacker task executes instructions on the core. %
  In the preparation phase, IPs in the SoC are %
  configured to ``spy'' on activities in the system.
\item \emph{Recording:} The second phase of the attack window begins with a context
  switch to the victim task. %
  The victim performs security-critical operations while, simultaneously, the spying IPs collect %
  information about the execution and store it in the system state. %
\item \emph{Retrieval:} In the third phase, control of the core is switched back to the
  attacker task. %
  The attacker retrieves the confidential information stored in the state of
  the system IPs. %
\end{itemize}
  
\begin{figure}
    \centering
    \includegraphics[trim={32mm 30mm 32mm 30mm}, clip, width=\linewidth]{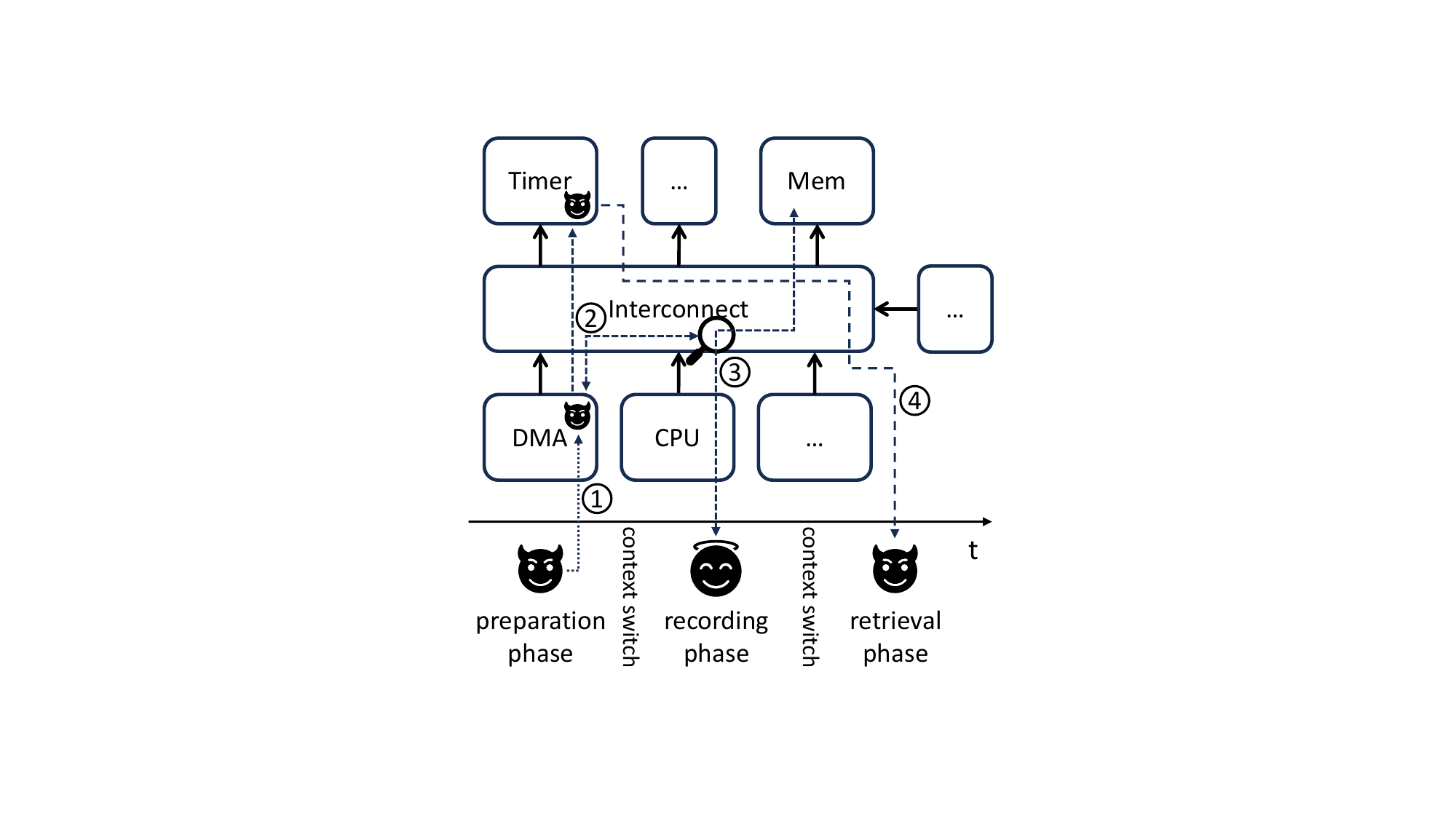} 
    \vspace{-5mm}
    \caption{Example of timing side channel attack on an MCU}
    \label{fig:attack-structure}
    \vspace{-5mm}
\end{figure}

As a simple example, consider an MCU SoC featuring,
amongst other IPs, a simple CPU, memory, a DMA and a timer,
as illustrated in Fig.~\ref{fig:attack-structure}. %
In the preparation phase, the attacker instructs the DMA to perform some
memory accesses and afterwards start the timer~\textcircled{1}. %
After a context switch to the victim task, the
recording phase of the attack begins. %
As instructed by the attacker, the DMA performs the
memory accesses and then starts the timer~\textcircled{2}. %
If the victim task tries to access the memory concurrently with the
DMA, contention is created~\textcircled{3}. %
As a result, the start of the timer is delayed.
In the retrieval phase of the attack, the attacker
task reads the timer state or waits for a timer overflow event~\textcircled{4}. %
Based on the timer state, the attacker can deduce information about
the victim task's memory accesses.
This information can then be used to break the isolation provided by the security infrastructure,
as previous attacks have demonstrated~\CITE{2024-RodriguesOliveira.etal}. %

\section{Formal detection method}
\label{sec:method}

\subsection{Basic Idea}

This work presents a novel formal verification method for detecting
vulnerabilities to SoC-wide timing side channel attacks, as described in Sec.~\ref{sec:model}. %
The method builds upon an existing security verification technique called
Unique Program Execution Checking (UPEC)~\CITE{2023-FadihehWezel.etal}. %
One key ingredient of UPEC is the underlying computational model which is a \emph{bounded} one,
i.e., its properties describe finite time windows. %
However, it employs a symbolic starting state and thus generates proof results with \textit{unbounded validity}. %
Nonetheless, while it is, in principle,
possible to describe the entire time window in which an attack occurs in a finite UPEC property,
this entails hundreds, if not thousands of clock cycles of behavior. %
Verifying such properties for realistic MCU designs is computationally infeasible. %

Fortunately, it is not necessary to capture the entire three-phase attack sequence described above in a
UPEC property, as we now explain. %
  
\begin{observation}
    \label{obs:victim-phase}
		All victim behavior, which is considered confidential per our threat model, can be spied upon in the recording phase only after the victim task accesses memory or peripheral devices. In the HW implementation these accesses are sent and received at the interface between the CPU and the rest of the system. 
\end{observation}

Understanding the implications of this observation requires a closer look at the system. %
We inspect the design under verification at the Register Transfer Level (RTL). %
This allows us to reason in terms of the state variables, i.e., all state-holding elements (incl. memory), of the design. %
We denote the set of all state variables as $S_\textit{all}$ and define the following subset: %
  
\begin{definition}
    \label{def:s-not-victim}
    The set $S_{\lnot\textit{victim}} \subset S_\textit{all}$ contains
    all state variables in the system except for the ones belonging %
    \begin{enumerate}
    \item to the CPU or
    \item to the memory locations allocated to the victim task.
    \end{enumerate}
\end{definition}

\begin{figure}
    \centering
    \includegraphics[trim={42mm 18mm 42mm 103mm}, clip, width=\linewidth]{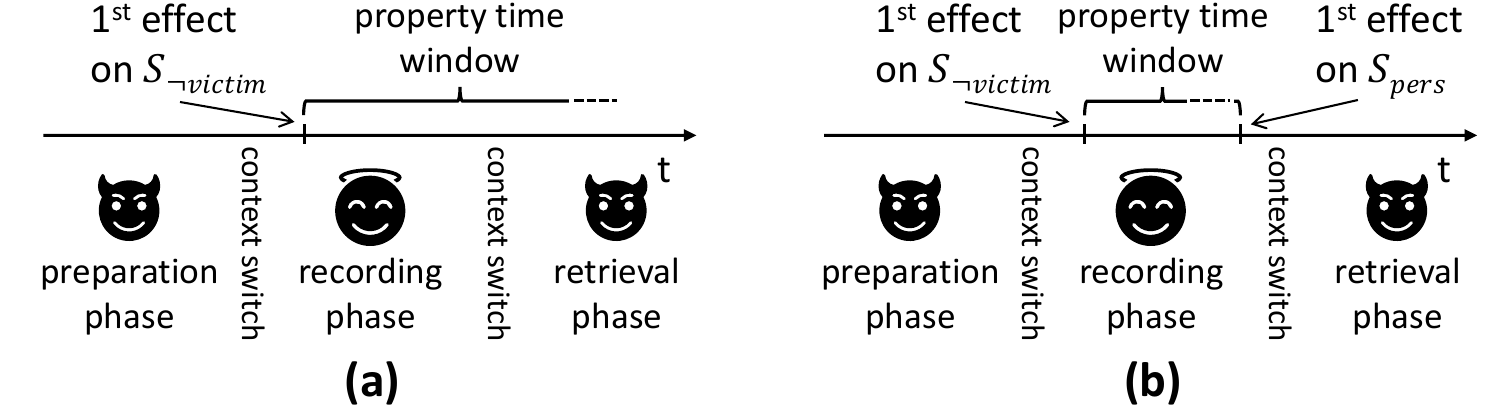} 
    \vspace{-7mm}
    \caption{Reduced property time window, (a)~after applying Obs.~1, and~(b)~after applying Obs.~1 and Obs.~2}
    \vspace{-3mm}
    \label{fig:timeline}
\end{figure}

Specifying $S_{\lnot\textit{victim}}$ is usually straightforward and is
achieved by simple structural analysis of the RTL model and by
symbolic modeling of memory address ranges. %
We provide more details on this process in Sec.~\ref{sec:scope}. %
Following Obs.~\ref{obs:victim-phase}, it becomes clear that spying
on the victim is possible only if at least one state variable
of~$S_{\lnot\textit{victim}}$, such as a control register of the interconnect,
is affected during the recording
phase by a dependency on the victim's confidential data. %
This helps us to reduce the property time window by excluding the
preparation phase, as illustrated in Fig.~\ref{fig:timeline}~(a). %
Using a symbolic starting state,
the time window starts when the victim execution affects the state
variables in $S_{\lnot\textit{victim}}$ for the first time, during %
the recording phase.

We thereby decrease the number of clock cycles the property has to
cover while still capturing any spying on security-critical victim
behavior. %
However, describing the retrieval phase in the property,
i.e., the information retrieval from the spying IPs, may require
the property to span some
additional hundreds of clock cycles, especially because %
this phase includes the context switch from victim to attacker. %
A second observation helps us to cope with this issue: 

\begin{observation}
  \label{obs:s-pers}
    For confidential information to be retrievable by an attacker
    task, it must persist in state variables in the system across a
    context switch.  %
\end{observation}

\begin{definition}
\label{def:s-pers}
We define $S_\textit{pers} \subset S_{\lnot\textit{victim}}$ as the set of state variables
\begin{enumerate}
\item that are accessible to the attacker task, and
\item that are \textit{persistent} across a context switch from the attacker to the victim task.
\end{enumerate} 
\end{definition}

Any attacker exploitation requires confidential information to end up in
$S_\textit{pers}$ at the end of the recording phase. %
With this observation the property time window can be further reduced
because we only need to prove that the victim execution
cannot affect any persistent state variables $S_\textit{pers}$ in the design.
As illustrated in Fig.~\ref{fig:timeline}~(b), this %
gives us a definitive end point of the property time window. %
We provide details on how to obtain $S_\textit{pers}$ in
Sec.~\ref{sec:scope} and how to prove by induction that the end point is valid. %

In the remaining parts of this section, we sketch the UPEC computational
model (Sec.~\ref{sec:background}) and then concretize the above observations
to formulate and prove bounded UPEC properties that span only a small number
of clock cycles while achieving global,
unbounded proof validity (Sec.~\ref{sec:method-core}, \ref{sec:scope}
and~\ref{sec:unrolled}).

\subsection{Background: UPEC computational model}
\label{sec:background}

UPEC was originally introduced as an exhaustive detection method for
transient execution attacks in CPUs. %
However, recent research has demonstrated its adaptability and its successful application to other
security verification problems~\CITE{2021-MuellerFadiheh.etal, 2022-DeutschmannMueller.etal, 2023-MehmedagicFadiheh.etal}. %
UPEC is based on Interval Property Checking
(IPC)~\CITE{2014-UrdahlStoffel.etal} at the RTL. %
In IPC, properties are formulated over a finite number of clock cycles
using the RTL design's state variables and signals. %
For checking a property, the IPC solver employs a symbolic starting
state %
which models all possible histories of inputs to the design. %
This is different from bounded model checking, where a concrete starting state is used. %
 
The goal of a UPEC proof is to guarantee that an attacker task cannot
extract confidential data located in the memory. %
This is achieved by a special 2-safety computational model: %
The RTL design is instantiated twice and all state variables of the
design are assumed to be equal between the two instances, except for
the confidential data. %
The UPEC property assumes an attacker task to execute on the processor
core and proves \textit{uniqueness} of the task execution, %
i.e., independence of the execution on the confidential data. %
This is checked by verifying that any difference between the two
instances, which must originate from confidential data, cannot propagate to the attacker task and influence its execution. %

\subsection{The UPEC-SSC method}
\label{sec:method-core}

We exploit Obs.~\ref{obs:victim-phase}~and~\ref{obs:s-pers}, %
as described above, and formulate a UPEC property spanning only two clock cycles,
as depicted in Fig.~\ref{fig:property}. The property asserts the macro
\textit{Primary\_Input\_Constraints()} which constrains the primary
inputs to be equal between the two instances. 
With the macro \textit{Victim\_Task\_Executing()} the property
assumes that all address ranges representing memory regions or
device registers allocated to the victim task are not directly
accessible by potentially spying IPs in the SoC.
This restricts the property to our threat model.
In addition, the macro constrains victim accesses to protected
addresses to be different between the two system instances
while accesses to all other addresses are equal.  %
Thus, only protected accesses are modeled as confidential information. %
The property makes no restrictions regarding the actual program executed
as victim task; any possible software is considered by the IPC solver. %
Furthermore, the address ranges allocated to the victim task are modeled
symbolically to represent any possible allocation. %
The macro \textit{State\_Equivalence($S$)} is parameterized with %
a set of state variables~$S$. It requires all state variables in~$S$ to be
equal to their counterparts in the other SoC instance. %
We use~$S$ to model the part of the system that has not been affected
by the victim behavior and thus cannot hold confidential information
at clock cycle~$t$. %
The property proves that also in clock cycle~$t+1$ the unaffected part
of the system remains unaffected by the confidential victim
information. %

\begin{figure}
  \centering
  \begin{minipage}{0.9\linewidth}
      \fontsize{9}{11}\selectfont
      \begin{tabbing}
      XX\=XXXXXX\=XXX\=XXXX\=\kill
      \textbf{UPEC-SSC($S$)}:\\
      \textcolor{blue}{assume:} \\
      \> \textcolor{blue}{during} $t$..$t$+$1$: \>\>\>\textit{Victim\_Task\_Executing()}; \\
      \> \textcolor{blue}{during} $t$..$t$+$1$: \>\>\>\textit{Primary\_Input\_Constraints()}; \\
      \> \textcolor{blue}{at} $t$: \>\>\>\textit{State\_Equivalence($S$)}; \\
      \textcolor{blue}{prove:} \\
      \> \textcolor{blue}{at} $t$+$1$: \>\>\>\textit{State\_Equivalence($S$)}; \\
      \end{tabbing}
  \end{minipage}
  \vspace{-5mm}
  \caption{2-cycle UPEC-SCC property}
  \vspace{-2mm}
  \label{fig:property}
\end{figure}
  
This property is the centerpiece of the proposed
\textit{UPEC for System Side Channels (UPEC-SSC)} method. %
The verification procedure is sketched in \refalg{upec-soc}. %
It starts with initializing~$S$ with
$S_{\lnot\textit{victim}}$, i.e., the system
state that is unaffected by the victim task execution before the
victim accesses any component outside the core. %
Then, the UPEC-SSC property is checked for the state set~$S$ using an
IPC solver and a counterexample is produced. %
The counterexample shows a violation of the property, i.e., a set of
state variables $S_\textit{cex}$ that are different from
their counterparts in the other design instance. %
We distinguish three cases for $S_\textit{cex}$: %
In the first case, the property holds and
$S_\textit{cex} = \emptyset$. %
This means the victim cannot influence additional state variables, the
design is secure, and the procedure terminates. %
In the second case, the property fails and $S_\textit{cex}$ includes
elements from $S_\textit{pers}$, i.e., data about the victim execution
can propagate to a persistent IP, where it can be retrieved later by
the attacker task. %
This case constitutes a vulnerability to an attack. %
In the third case, the property fails, but $S_\textit{cex}$ does
not include elements from $S_\textit{pers}$. %
In this case, the victim execution influences new state variables, but
they cannot hold secret data across a context switch to the attacker task. %
As a result, the state variables in $S_\textit{cex}$ are removed from
the non-influenced set of states~$S$ for the next check of the property. %
In every iteration of the \textit{while} loop, the set of states
variables~$S$ to which secret information from the victim can
possibly propagate, is reduced. %
This is repeated until, eventually, %
the design is either proven secure or
vulnerable. %

For a secure design, the procedure in \refalg{upec-soc} computes a
set of state variables, $S$, for which property UPEC-SSC($S$)
becomes inductive. %
In the final iteration that leads to return \textit{secure}
(line~6), $S$ has reached a fixed point in which property
UPEC-SSC($S$) (called in line~4) becomes the \emph{step} of the
induction. %
It proves, for the final set~$S$, that if the victim has not
influenced any state variable in~$S$ at a time~$t$, it will also not
do so at any time in the future. %
The induction \emph{base} is given trivially by the fact that
initially, before the victim becomes active on the CPU/system
interface, no state variable has been influenced at all. %
For the final set $S$ it holds that $S_\textit{pers} \subset S \subset S_{\lnot\textit{victim}}$.
Hence, it is proven that the victim has not caused any change of the system state that
could be visible to the attacker after a context switch.

\begin{algorithm}
    \caption{UPEC-SSC Procedure}
    \label{alg:upec-soc}
    \begin{algorithmic}[1]
      \Procedure{UPEC-SSC}{}
      \State $S \gets S_{\lnot\textit{victim}}$
      \While {\textit{true}}
      \State $S_\textit{cex} \gets $ check(UPEC-SSC($S$))
      \If {$S_\textit{cex} = \emptyset$}
      \State \Return \textit{secure}
      \ElsIf {$S_\textit{cex} \cap S_\textit{pers} \neq \emptyset$}
      \State \Return (\textit{vulnerable}, $S_\textit{cex}$)
      \ElsIf {$S_\textit{cex} \cap S_\textit{pers} = \emptyset$}
      \State $S \gets  S \setminus S_\textit{cex}$
      \EndIf
      \EndWhile
      \EndProcedure
    \end{algorithmic}
\end{algorithm}

\vspace{-3mm}

\subsection{UPEC-SCC state variables}
\label{sec:scope}

The UPEC-SSC procedure proves non-interference of the victim task with the persistent states in the SoC, under the given constraints
and for an arbitrary behavior of the system IPs. %
\par The procedure %
requires the specification of the two state variable sets
$S_{\lnot\textit{victim}}$ and $S_\textit{pers}$. %
The set $S_{\lnot\textit{victim}} \subset S_\textit{all}$ is compiled by excluding all state variables in the core and
the victim memory space, from the set of all state variables in the SoC. %
Obtaining the state variables in the core is
straightforward and requires simple structural analysis with minimal manual input from a verification engineer. %
The victim memory space, i.e., the memory regions containing the victim program and data,
are determined by address ranges in the memory devices of the SoCs. %
We model the address ranges symbolically. %
This includes all possible victim memory layouts into the properties
and allows the proofs to be independent of any particular victim software. %
The set of all state variables that are persistent, $S_\textit{pers}$, is more challenging to compute. %
Determining this set for the entire SoC might be infeasible. %
This is, however, not needed. %
The procedure only requires us to specify whether state variables
are part of $S_\textit{pers}$ if they appear in a counterexample. %
In practice, for the considered property, most of these state variables are buffers
in the interconnect which are overwritten with every communication
transaction and cannot be used to hold persistent information. %
Therefore, they are not part of $S_\textit{pers}$. %
The remaining state variables included in counterexamples are
usually registers or memory arrays in IPs and must be considered persistent. %
Some rare counterexamples may involve state variables that are neither buffers in the interconnect nor obviously
persistent registers in IPs. %
These cases require closer inspection and additional proofs
to qualify or disqualify them as part of $S_\textit{pers}$. %
However, in all our experiments we have not observed any such cases. %
In fact, conservative fixes can block propagation to state variables beyond the interconnect,
as discussed in Sec.~\ref{sec:countermeasure}.

IPC properties, such as the employed UPEC-SSC property from Fig.~\ref{fig:property}, can produce false counterexamples. %
This happens when the solver assumes an unreachable state as its
starting state. %
The standard solution in IPC %
is to employ invariants that exclude the unreachable behavior from the symbolic starting state. %
In some applications of IPC, formulating and proving invariants
may require significant manual effort. %
The false counterexamples encountered in our experiments, however, involve 
only few state variables and the associated invariants are straightforward to formulate. %

\subsection{Generating longer counterexamples}
\label{sec:unrolled}

The UPEC-SSC property in Fig.~\ref{fig:property} proves a design under
verification to be either secure or vulnerable within two clock
cycles. %
Counterexamples to the property also span two clock cycles. %
Behavior affecting multiple state variables at multiple time points is
included \textit{implicitly} in the computed starting state of the counterexamples. %
In many cases, such implicit representations of behavior are cryptic
and the corresponding counterexamples
are of limited use to a verification engineer. %
Therefore, it is desirable to obtain longer
counterexamples containing all signal valuations \textit{explicitly}. %
Such counterexamples can be generated by unrolling the property over
more than two clock cycles. %
Compared to the two-clock-cycle version from Fig.~\ref{fig:property},
the unrolled property, described in Fig.~\ref{fig:unrolled-property},
has an additional parameter~$k$ that %
controls the number of unrollings. %
Secondly, the property uses a vector of state sets, %
~$S[$~$]$, to specify for each clock cycle a set of states which are %
not influenced by the victim task at this particular clock cycle. %

\begin{figure}
\centering
\begin{minipage}{0.9\linewidth}
    \fontsize{9}{11}\selectfont
    \begin{tabbing}
    XX\=XXXXXX\=XXX\=XXXX\=\kill
    \textbf{UPEC-SSC-unrolled($k, S[$ $]$)}:\\
    \textcolor{blue}{assume:} \\
    \> \textcolor{blue}{during} $t$..$t$+$1$: \>\>\>\textit{Victim\_Task\_Executing()}; \\
    \> \textcolor{blue}{during} $t$..$t$+$k$: \>\>\>\textit{Primary\_Input\_Constraints()}; \\
    \> \textcolor{blue}{at} $t$: \>\>\>\textit{State\_Equivalence($S[0]$)}; \\
    \textcolor{blue}{prove:} \\
    \> \textcolor{blue}{at} $t$+$1$: \>\>\>\textit{State\_Equivalence($S[1]$)}; \\
    \> ... \\
    \> \textcolor{blue}{at} $t$+$k$: \>\>\>\textit{State\_Equivalence($S[k]$)}; \\
    \end{tabbing}
\end{minipage}
\vspace{-5mm}
\caption{Multi-cycle UPEC-SCC property}
\vspace{-2mm}
\label{fig:unrolled-property}
\end{figure}

The unrolled property leads us to the unrolled UPEC-SSC procedure~\refalg{upec-soc-unrolled}. %
It starts by initializing the two sets $S[0]$ and $S[1]$ with $S_{\lnot\textit{victim}}$ %
and the parameter~$k$ with~$1$. %
The property is then checked with the initialized parameters. %
Note that this configuration of the property is identical to the first
iteration in the original procedure in \refalg{upec-soc}. %
If the property fails due to a difference in a persistent state, a
vulnerability is detected. %
For all other counterexamples, the diverging state variables are
removed from the set of affected state variables $S[1]$, for the next
run of the property check. %
However, compared to the original procedure, $S[0]$, i.e., the victim
task's influence at the beginning of the time window~$t$, remains the same in all iterations. %
Checking the property and adjusting $S[1]$ is repeated until no new
counterexamples are detected and the property holds. %
Then, the property is unrolled an additional clock cycle by
incrementing~$k$ and the set~$S$ is initialized with the set of the
previous clock cycle. %
In this way, the property is unrolled clock cycle by clock cycle,
until the victim task cannot influence %
any new state variables, i.e., the
property holds for a new unrolling~$k$ without removing state variables
from~$S[k]$, or a vulnerability is detected. %

\begin{algorithm}
\caption{Unrolled UPEC-SSC Procedure}
\label{alg:upec-soc-unrolled}
\begin{algorithmic}[1]
    \Procedure{UPEC-SSC-UNROLLED}{}
    \State
    $S[0], S[1] \gets S_{\lnot\textit{victim}}, S_{\lnot\textit{victim}}$
    \State $k \gets 1$ 
    \While {\textit{true}}
    \State $S_\textit{cex}$ $\gets $ check(UPEC-SSC-unrolled($k, S$))
    \If {$S_\textit{cex} = \emptyset$}  
    \If {$S[k] = S[k-1]$} 
    \State \Return \textit{hold}
    \Else
    \State $k \gets k + 1$
    \State $S[k] \gets S[k-1]$
    \EndIf
    \ElsIf {$S_\textit{cex} \cap S_\textit{pers} \neq \emptyset$}
    \State \Return (\textit{vulnerable}, $S_\textit{cex}$)
    \ElsIf {$S_\textit{cex} \cap S_\textit{pers} = \emptyset$}
    \State $S[k] \gets  S[k] \setminus S_\textit{cex}$
    \EndIf
    \EndWhile
    \EndProcedure
    \end{algorithmic}
\end{algorithm}
\vspace{-2mm}

Note that, even if the procedure returns \emph{hold},
an additional inductive proof is required to prove the system secure. %
This is because, even if the victim task's influence could not propagate
to new state variables from clock cycle~$k-1$ to~$k$,
there might be a future clock cycle~$k+n$ where the propagation continues. %
Performing the inductive proof can be done
by using the original algorithm (Alg.~\ref{alg:upec-soc}) with $S \gets S[k]$
from the last unrolled clock cycle~$k$. %

\section{Case study}
\label{sec:experiments}

We conducted a case study on Pulpissimo~\CITE{2018-SchiavoneRossi.etal} which is an MCU-style SoC featuring a 2-stage pipelined \RISCV{} core, a DMA,
an accelerator, and many peripheral IPs, including I/O interfaces.
The entire SoC comprises more that 5M state variables. %
Applying UPEC-SCC to Pulpissimo yielded several counterexamples
which point to potential vulnerabilities in the MCU. 
We started our analysis and debugging process by picking a counterexample
that is of particular interest because it points to
a so far unknown variant of the BUSted attack. %
In the following, we provide insights into this new attack variant and %
the detection of the corresponding vulnerability in Sec.~\ref{sec:attack}. %
In Sec.~\ref{sec:countermeasure}, we elaborate on how to mitigate such vulnerabilities
and how the UPEC-SSC method was used to prove the modified
system secure w.r.t.\ the security objectives. %


We conducted all experiments on a workstation PC featuring an Intel
i9-13900k with 128 GB RAM running Linux and the commercial property
checker OneSpin 360 DV by Siemens EDA. %

\subsection{New BUSted variant: accelerator + memory}
\label{sec:attack}

UPEC-SCC generated a counterexample revealing a security vulnerability involving
the Hardware Performance Engine (HWPE) accelerator IP
and a memory device in the system. %
The HWPE can be configured to fetch its inputs directly from the memory, perform
complex arithmetic operations on the data, and write the results %
back to a configured memory region. %

The detected counterexample shows a scenario where the attacker sets
up the HWPE to spy, i.e., to collect information about the victim task's memory accesses. %
Following the attack structure introduced in Sec.~\ref{sec:attack-structure},
the scenario can be divided into three phases. %
In the preparation phase, the attacker task chooses a memory region with write
permissions and primes it with zeros. %
It then sets up the HWPE accelerator IP to progressively overwrite the
memory region with non-zero values. %
The recording phase starts after a context switch to the victim task. %
The victim accesses a different memory location in the same memory
device, concurrently with the HWPE's memory transactions. %
This behavior creates contention on the interconnect and the
continuous accesses by the HWPE are delayed. %
In the retrieval phase of the attack,
control in the core is once again transferred %
to the attacker task. %
The primed memory region contains information about
the victim's memory accesses: %
By examining the progress made by the HWPE %
in overwriting the primed region, the attacker can deduce the number
of memory accesses the victim has made to a particular memory device
and launch an attack similarly as in the original BUSted attack~\CITE{2024-RodriguesOliveira.etal}.

We detected this scenario with the unrolled %
proof procedure of Alg.~\ref{alg:upec-soc-unrolled}.
The preparation phase of the attack is implicitly contained in the symbolic initial state of the proof. %
The proof window starts in the recording phase,
when the victim task accesses the memory for the first time. %
We unrolled for 2~clock cycles to observe the delay of the HWPE memory access. %
The retrieval phase is implicitly modeled by classifying the state variables of
the attacker-accessible memory region as persistent state variables and part of $S_\textit{pers}$. %
The runtime of the proof iterations was below one minute. %

The detected vulnerability has an important characteristic:
It allows an attacker to open a timing side channel without the use of an actual timer.
This undermines a cheap and popular countermeasure
against timing attacks, where access to system timers is denied
to untrusted tasks or artificial noise is added to timers to obstruct
precise time measurement~\CITE{1991-Hu}.  

\vspace{-4mm}
\subsection{Countermeasure}
\label{sec:countermeasure}

For the purpose of evaluating UPEC-SCC in the case of a secure design,
we implemented a conservative countermeasure that fixes %
the vulnerability described above. %
The fix could be proven to also eliminate all other counterexamples. %
We designated a memory region for the security-critical operation of the victim task and
restricted the access to the corresponding memory device. %
We make use of Pulpissimo's architecture that features two types of memory,
a public and a private memory,
which are implemented as separate memory devices. %
The two memory devices are connected to the SoC by two separate crossbars. %
The private memory is only accessible to a few IPs (including the processor core). %
We mapped the security-critical memory region into the address space of the private memory device. %
Consequently, we only needed to restrict the access for very few IPs. %
The restrictions consist of a set of legal configurations for the corresponding IPs and %
can be compiled as a set of firmware constraints to be checked for compliance during firmware development and verification. %
This is similar to the process introduced in~\CITE{2021-MuellerFadiheh.etal}. %

With this countermeasure in place, we ran the proof procedure of Alg.~\ref{alg:upec-soc}.
After~3 iterations, the procedure proved the system to be secure w.r.t.\ the considered threat model. %
The runtime of the iterations ranged between 58~seconds and 2~hours~52~minutes. %

\section{Related Work}
\label{sec:related-work}

Only little research has been conducted on timing side channel attacks
in MCU-style SoCs with single processors. %
BUSted~\CITE{2024-RodriguesOliveira.etal}
is the first comprehensive attack that uses MCU peripheral IPs to spy on %
a running victim task.
The attack leverages contention on the MCU bus %
to breach Armv8-M TrustZone and retrieve confidential data.
A similar timing side channel involving a DMA creating contention with a victim execution
has been presented in~\CITE{2022-BognarVanBulck.etal}. 
This and the BUSted attacks are good examples for the type of vulnerabilities
our presented method detects. In a Nemesis attack~\CITE{2018-VanBulckPiessens.etal},
instruction-dependent timing differences in the CPU interrupt logic are exploited
to breach the isolation of the Sancus Trust Execution Environment (TEE).
Even though Nemesis is demonstrated on an MCU,
the exploited side channel (instruction execution time) is
located inside the CPU and is therefore not within the class of vulnerabilities
considered in this work.

There are several related types of timing attacks targeting systems
that allow attacker concurrency or include caches.
Contention-based attacks (e.g.,~\CITE{2022-ZhaoMorrison.etal, 2014-WangFerraiuolo.etal})
exploit simultaneous accesses to shared caches or memories in multi-core or multi-threaded CPUs. %
Timing attacks in NoCs (e.g.,~\CITE{2015-SepulvedaDiguet.etal, 2016-ReinbrechtSusin.etal})
deduce confidential information by accessing shared routers,
while attacker and victim tasks are running on different network nodes.
In cache-based timing side channels (e.g.,~\CITE{2005-Percival, 2014-YaromFalkner})
footprints left by a victim in caches produce differences in hit and miss times and
allow an attacker to deduce confidential data.
All these attacks are well understood and a plethora of effective countermeasures
has been proposed.
We therefore concentrate on the relatively new class of MCU-wide timing side channels,
which pose a severe problem even without exploiting attacker concurrency or
buffers like caches.

There are no dedicated formal methods for detecting
vulnerabilities in MCUs. %
Formal verification methods targeting related security problems in
similar systems are usually not well suited for exhaustive detection
of timing side channels in MCU-style SoCs. %
Property checking approaches like~\CITE{2019-FarzanaRahman.etal}
consider individual security properties targeting a limited set of
security problems in SoCs. %
They do not focus on providing SoC-wide guarantees. %
Another popular verification approach applied to security problems in SoCs is Information Flow Tracking (IFT). %
As summarized in~\CITE{2022-HuArdeshiricham.etal}, IFT computes the information flow between a designated pair of source and sink in a design. %
We believe that, in principle, detecting timing side channels in SoCs can be formulated as an IFT problem. %
However, exhaustively covering \textit{all} possible vulnerabilities seems unattainable. %
In addition, it is unlikely that contemporary solvers can handle the complexity of
global IFT analysis in SoCs of realistic size. %
There are two works that apply exhaustive formal verification for detecting security vulnerabilities in SoCs. %
In~\CITE{2021-MuellerFadiheh.etal} a 2-safety property is used to detect all functional design bugs causing confidentiality violations
in SoCs. %
The work in~\CITE{2023-MehmedagicFadiheh.etal} exhaustively detects integrity caused by untrusted third-party IPs in SoCs. %
Both approaches can guarantee the absence of vulnerabilities w.r.t.\ their respective security objectives. %
But, unlike this work, they do not target and do not scale to SoC-wide timing side channels. %

\section{Conclusion}
\label{sec:conclusion}

In this paper, we present a novel formal verification methodology for
detecting timing side channels in MCU-style SoCs. %
The method detected a so far unknown variant of a timing side channel
attack which allows the attacker to record information about a victim
task's memory accesses without requiring cache footprints or
access to hardware timers. %
We demonstrate that the method is scalable for an SoC of realistic
size and can prove the SoC to be secure w.r.t.\ the security objective
after we implemented a conservative countermeasure. %
Our future work will explore a UPEC-SCC driven design methodology leading
to new and less conservative countermeasures. %

\bibliographystyle{ACM-Reference-Format}
\bibliography{refs3}

\end{document}